# Boosting Thermoelectric Power Factor of Carbon Nanotube Networks with Excluded Volume by Co-embedded Microparticles

Oluwasegun Isaac Akinboye[1], Yu Zhang[1], Vamsi Krishna Reddy Kondapalli[1], Fan Yang[2], Vesselin Shanov[3], Yue Wu[2], and Je-Hyeong Bahk[1*]

[1]Department of Mechanical and Materials Engineering, University of Cincinnati, Cincinnati, OH 45221, USA

[2] Department of Chemical and Biological Engineering, Iowa State University, Ames, IA 50011, USA

[3] Department of Chemical and Environmental Engineering, University of Cincinnati, Cincinnati, OH 45221, USA

*Email: bahkjg@ucmail.uc.edu

**ABSTRACT:**

Carbon nanotube (CNTs) networks embedded in polymer matrix have been extensively studied over the recent years as a flexible thermoelectric (TE) transport medium. However, their power factor has been largely limited by the relatively inefficient tunneling transport at junctions between CNTs and the low density of conducting channels through the networks. In this work, we demonstrate that significant enhancement of power factor is possible by adding electrically insulating microscale particles in CNT networks. When silica particles of a few µm diameters were co-embedded in single-walled CNT-polydimethylsiloxane (PDMS) composites, both the electrical conductivity and the Seebeck coefficient were simultaneously enhanced, thereby boosting the power factor by more than a factor of six. We find that the silica microparticles excluded a large volume of the composite from the access of CNTs and caused CNT networks to form around them using the polymer as a binder, which in turn resulted in improved connectivity and alignment of CNTs. Our theoretical calculations based on junction tunneling transport show that the large power factor enhancement can be attributed to the enhanced tunneling with reduced junction distance between CNTs and the increased geometric factor due to better CNT alignment. Additional enhancement of power factor by more than a factor of two was achieved by sample compression due to the further improvement of CNT alignment.

## 1. Introduction

Research on thermoelectric (TE) materials has accelerated in recent years due to the world-wide efforts for low-grade waste heat recovery and wearable body-heat harvesting.[1,2] The performance of a TE material is evaluated using the unitless figure-of-merit, $ZT = S^2\sigma T/\kappa$, which is dependent on the Seebeck coefficient ($S$), electrical conductivity ($\sigma$), and thermal conductivity ($\kappa$).[3–5] $T$ is the absolute temperature. The numerator part ($S^2\sigma$) in $ZT$ is called the power factor (PF), which determines the power output of a TE device for a given temperature difference. PF is a material property determined by charge carrier transport in the material. The trade-off relationship between $S$, $\sigma$, and $\kappa$ is a fundamental issue that has hampered further enhancement of TE materials performance.[6–10]

Due to their excellent chemical stability, durable mechanical qualities, and outstanding electric properties, carbon nanotubes (CNTs) are frequently utilized as fillers in TE nanocomposites and have thus emerged as a promising choice for the construction of novel flexible thermoelectric devices. [6,11] Furthermore, with the sp$^2$-hybridized nature of CNTs, it is advantageous when considering TE applications for nanocarbon-based hybrid composites.[3,12–14] Recently, several reports explored the TE performances of individual CNTs and their hybrid composites and they were able to achieve a high PF at room temperature.[15–20] However, the TE properties of CNT-based composites have been limited due to inefficient tunneling transport at the junctions between CNTs.[12]

Recently, Park et al. [23] showed that the electrical conductivity of CNT networks could be enhanced by adding microscale particles as secondary fillers. These microparticles (MPs) created excluded volume and pinched CNTs into the narrow space between the microparticles to form denser, better-connected percolation networks of CNTs, resulting in enhanced electrical conductivity. The effect of the excluded volume on TE transport particularly on the Seebeck coefficient, however, has not been fully investigated. Furthermore, due to the inclusion of secondary fillers – silica MPs, which has the tendency to impede the interconnections between CNTs, previous investigations have seen irregular trends in electrical conductivity of the hybrid composite systems.[21–23]

In this work, we combine experimental and theoretical studies to investigate the effects of microparticles co-embedded with CNTs in a polymer matrix on the power factor. Polydimethylsiloxane (PDMS) is employed as the matrix for this study due to its many advantages, such as solution processability, lightweight, biocompatibility, and low thermal conductivity.[24] PDMS also works as an excellent binder between CNTs and silica particles. Since both PDMS and silica particles are electrically insulating, all the carrier transport occurs through CNT networks. Hence, this material combination provides an excellent model system to study the variation of power factor exclusively by modification to CNT networks.

## 2. Experiments

Single walled carbon nanotubes (SWCNT), silica MPs, PDMS, are the raw components for the CNT-based polymer composites material development, and chloroform was used as the solvent. Dow Corning supplied commercially available PDMS (Sluggard 184 Silicone Elastomer Base) for the synthesis of the CNT-based polymer composite. Tuball SWCNTs from Sigma Aldrich with an average diameter of 1.6 nm and a length of > 5 µm was utilized. The SWCNTs are a hybrid of both semiconducting and metallic CNTs, in roughly 1:2 ratio. Silica powders of 1 and 3 µm particle size were bought from Sigma Aldrich. The aforementioned raw components were all used as received.

For the composite synthesis, it is imperative that the two fillers (SWCNT and silica) disperse evenly in the PDMS matrix. Despite their comparable structures, poor distribution conditions of these two filler materials in the PMDS matrix can generate opposing or irregular trends in electrical conductivity.[23,25] Ultrasonication was employed to guarantee good mixing and dispersion of entangled CNTs and silica microparticles in PDMS. They were disseminated in chloroform using ultrasonic agitation at a frequency of 20 KHz initially. These processes resulted in the homogenous dispersion conditions of the fillers in PDMS. More detailed information about the synthesis processes is found in Supporting Information: Sec. 1. Synthesis of CNT-PDMS composites. Scanning electron microscopy (SEM) was used to examine the morphological and structural characteristics of the composites. The samples were also examined using Raman spectroscopy (Renishaw inVia, stimulated by a 514 nm Ar – ion laser with a laser spot size of approximately 1 µm$^2$).

## 3. Results and Discussion

Figure 1 displays the cross-sectional SEM images of the CNT-silica MP-PDMS composites revealing the morphologies of CNT networks after incorporating with different content of silica

MPs. Compared with the sample without silica (as seen in Fig. 1(a)), all other samples showed hair-like CNT bundles connecting the silica particles like bridges. CNTs are also seen closely bound on the surface of silica particles, creating intimate networks on the particle surfaces with PDMS as a binder. The latter proves the creation of excluded volume due to the interactions between silica particles and CNTs. Moreover, it can be observed that as the silica content increased, the excluded volume becomes more apparent and the interconnectivity between the CNT bundles become more evident, thus contributing to the effective increase in electrical conductivity, which is presented later in this paper.

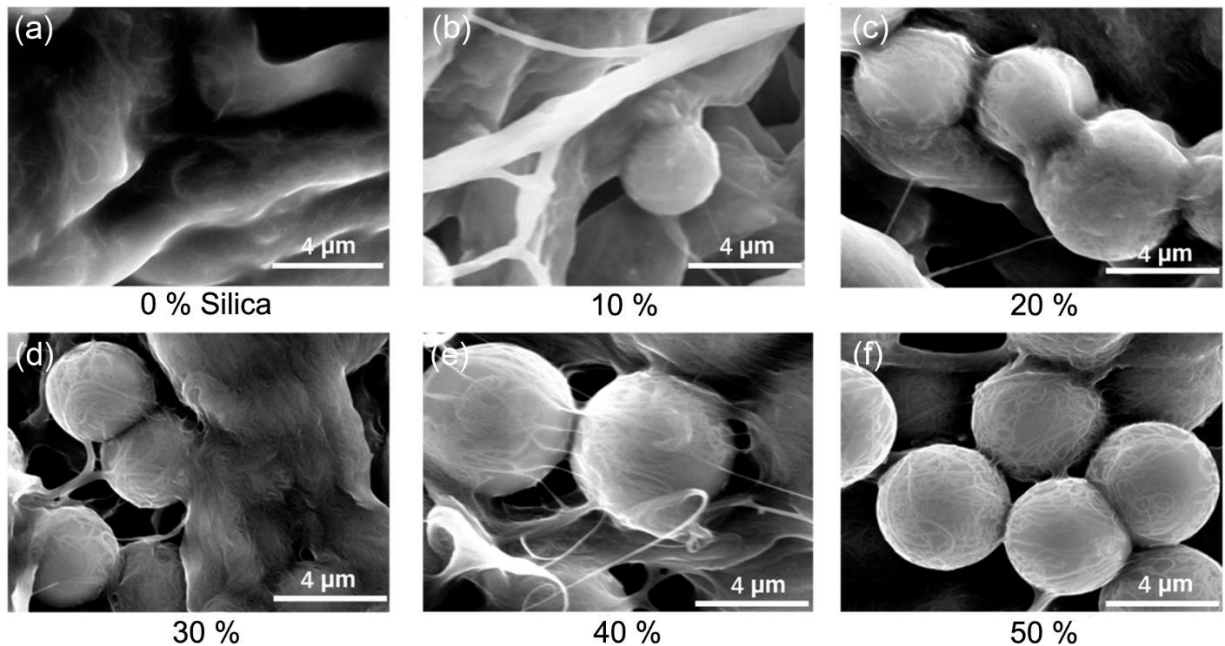

**Figure 1.** SEM images of SWCNT-silica MP-PDMS composites with varying content of silica particles (~3 µm diameter) from 0 to 50 w.t.%, and SWCNT content kept constant at 10 w.t.% PDMS content was varied accordingly, i.e., decreasing PDMS w.t.% with increasing silica content.

Fig. 2 shows the measured Seebeck coefficients of the composites as a function of silica content before and after compression. The method used for compression is discussed in Supporting Information: Sec. 7. Sample Compression. The values obtained from the experiment are all positive as shown in the figure, which is a clear indication that the majority carriers are holes. This is because the oxidation of CNTs generally makes them p-type in ambient environment. In both cases before and after compression, the Seebeck coefficient slightly decreased initially when a 10% silica MPs were added to CNT-PDMS composites. The initial small drop in Seebeck coefficient can be attributed to the disrupted CNT networks with addition of silica MPs, where the excluded volume effect was relatively too small to enhance the property. From 10 % to 40 % silica, the Seebeck coefficient steadily increased with increasing silica content. We employed the Landauer formalism for the junction tunneling transport between CNTs to explain this trend. More details about the theoretical calculation can be found in Ref. [12] and also in Supporting Information: Sec. 4. Thermoelectric transport calculations for CNT networks. Here, we used the average junction distance as a fitting parameter for both data before and after compression. Previously we found that the Seebeck coefficient can be enhanced when the junction distance decreases.[12] This is because the energy-dependent tunneling transmission ($\bar{T}$) increases more rapidly with energy ($E$) near the Fermi level, i.e. $\frac{d\bar{T}(E)}{dE}\Big|_{E=E_F}$ increases, as the junction distance increases, to ultimately promote the entropy transport by carriers. To the first-order approximation, we assumed a linear decrease of the junction distance with increasing silica content to fit the experimental Seebeck coefficients, which produced reasonable fitting for both before and after compression as shown in Fig. 2(a). Fig. 2(b) shows the average junction distance values used for the best curve fits as a function of silica content. Reduction of the average junction distance from ~ 12 to ~ 9 Å is obtained from the fitting for the as-synthesized samples before compression when 40 w.t.% silica is co-embedded. Further reduction of junction distance down to ~ 6 Å may be possible with sample compression as denser CNT networks can be created due to compaction.

Beyond 40 % silica, the decrease in Seebeck coefficient is observed as shown in Fig. 2(a), which can be attributed to the increased material inhomogeneity and more disruptive CNT networks with excess microparticles. As silica content increases, PDMS content is reduced accordingly. CNTs can still be bound well onto the surface of silica particles with less PDMS binder, but the hair-like inter-particle bridges made of PDMS-coated CNT bundles become apparently thinner and less observed due to the less amount of PDMS available as clearly seen in the SEM images in Fig. 1. We believe that these bridges play a critical role in connecting the sub-networks of CNTs on individual particles to create larger-scale conduction pathways from end to end of the sample. Increase silica content and reduced PDMS content seem to have made these inter-particle connections poorer, resulting in the reduction in the Seebeck coefficient. We model this behavior with increased distribution of junction distances while keeping the average value fixed. Below 40

% silica, we used a fixed standard deviation factor of 0.3, to model the standard deviation of junction distance ($\sigma_{std}$) that is directly proportional to the mean junction distance ($d_{mean}$), i.e., $\sigma_{std} = 0.3 d_{mean}$.[12] Beyond 40 % silica, we instead used an increasing standard deviation factor with increasing silica content from 0.3 to 1.2 for silica content from 40 % to 60 % to fit the experimental Seebeck coefficient reasonably well. (Dotted line in Fig. 2(a)) Note that samples with high silica contents above 40 % were relatively fragile due to the reduced PDMS content. These samples were damaged after compression; hence, the properties could not be measured.

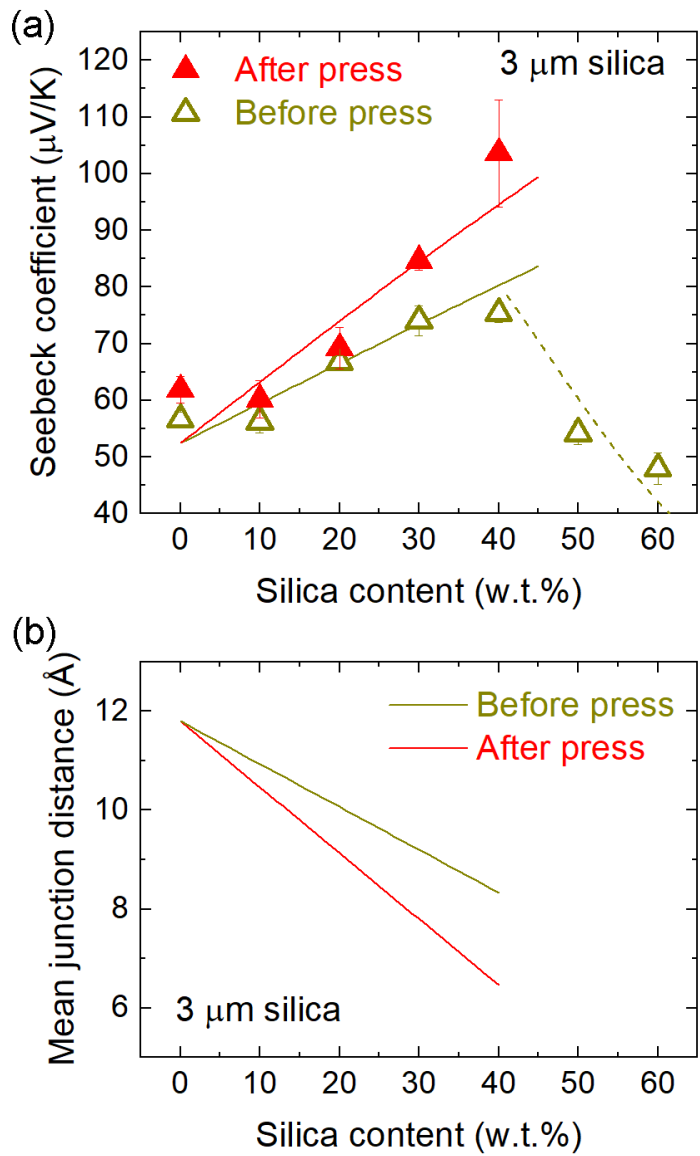

**Figure 2.** (a) Seebeck coefficient of SWCNT-silica MP-PDMS composites as a function of silica MP w.t.% for 3 μm particle size before and after compression. The content of SWCNT was fixed

at 10 w.t.%, and PDMS content was varied accordingly. Symbols are experimental data and curves are theoretical fitting. (b) Mean junction distances used in the theoretical fitting as a function of silica content up to 40 % silica. Beyond 40 % silica, the reduced Seebeck coefficient was fitted (the dotted line in (a)) with increasing standard deviation of junction distance with increasing silica content while the mean junction distance was kept constant to account for the increased material non-uniformity.

Fig. 3 shows the electrical conductivity of the same set of samples as a function of silica content. The electrical conductivity also increases with increasing silica content as the Seebeck coefficient does for both before and after compression. According to our tunneling transport theory, the electrical conductivity is given by

$$\sigma = \alpha G \qquad (1)$$

where $\alpha$ is the geometric factor in the unit of cm$^{-1}$ and $G$ is the average junction conductance in the unit of $\Omega^{-1}$. The geometric factor is defined as the ratio of the average density of independent conduction pathways per cross-sectional area ($m$) to the average number of junctions per length per conduction pathway in the current flow direction ($n$), such that $\alpha = \frac{m}{n}$. It is extremely difficult to determine the $m$ and $n$ experimentally as there is no good tool to investigate all the three-dimensional microstructure over a macroscopic scale. Instead, we use the geometric factor $\alpha$ as a fitting parameter to fit the experimental electrical conductivity data. The junction conductance $G$ is directly obtained from the junction characteristics such as the mean distance and potential height used to fit the Seebeck coefficient. So, we add only one additional fitting parameter here. Fig. 3(b) and (c) show, respectively, the junction conductance and the geometric factor used to fit the experimental electrical conductivity as a function of silica content. As the junction distance decreases with increasing silica content, the junction conductance is enhanced, which is the main cause of the electrical conductivity enhancement. The geometric factor, however, increases initially at low silica content, but slows down in the increase and eventually decreases with increasing silica content above 20 ~ 30 % silica. We believe this is possible due to the re-organization of CNT networks. Initially, better-connected CNT networks can be created by the addition of silica particles, so that more parallel conduction channels can be effectively created to increase *m*, and thus $\alpha$. As silica is added more and more, however, the same number of CNTs are re-organized and get closer to each other with reduced distance, which can make existing channels partially merged instead of making new ones, resulting in fewer number of channels. Hence, *m* and thus $\alpha$ could be reduced.

As shown in Fig. 3(a), the compression effect in conjunction with the creation of excluded volume led to a drastic increase, approximately 40% in the electrical conductivity of the CNT-PDMS composite specifically for high silica contents, 30 and 40 w.t.%. This enhancement largely comes from the reduced junction distances by compression, i.e., *G* increased by compression. At the same time, the geometric factor is also increased particularly at low silica content by compression, which indicates the improvement in the CNT network alignment. Since the samples were pressed down vertically, CNTs could be better aligned in the horizontal direction, which is the current direction for these in-plane measurements. After 20 ~ 30 % silica, however, the geometric factor after compression becomes smaller than that before compression, which might be indicative of increased number of junctions per length, i.e., increased *n*, along the horizontal direction. It should also be noted that irrespective of the silica size, the electrical conductivity values were somewhat similar considering the cases of either before compression or after compression. See Supporting Information: Sec. 5. TE properties of composites with 1 and 3 μm silica particles. This can be due to the fact that the density of conduction paths created by the excluded volume are similar for both 1 and 3 μm particle size.

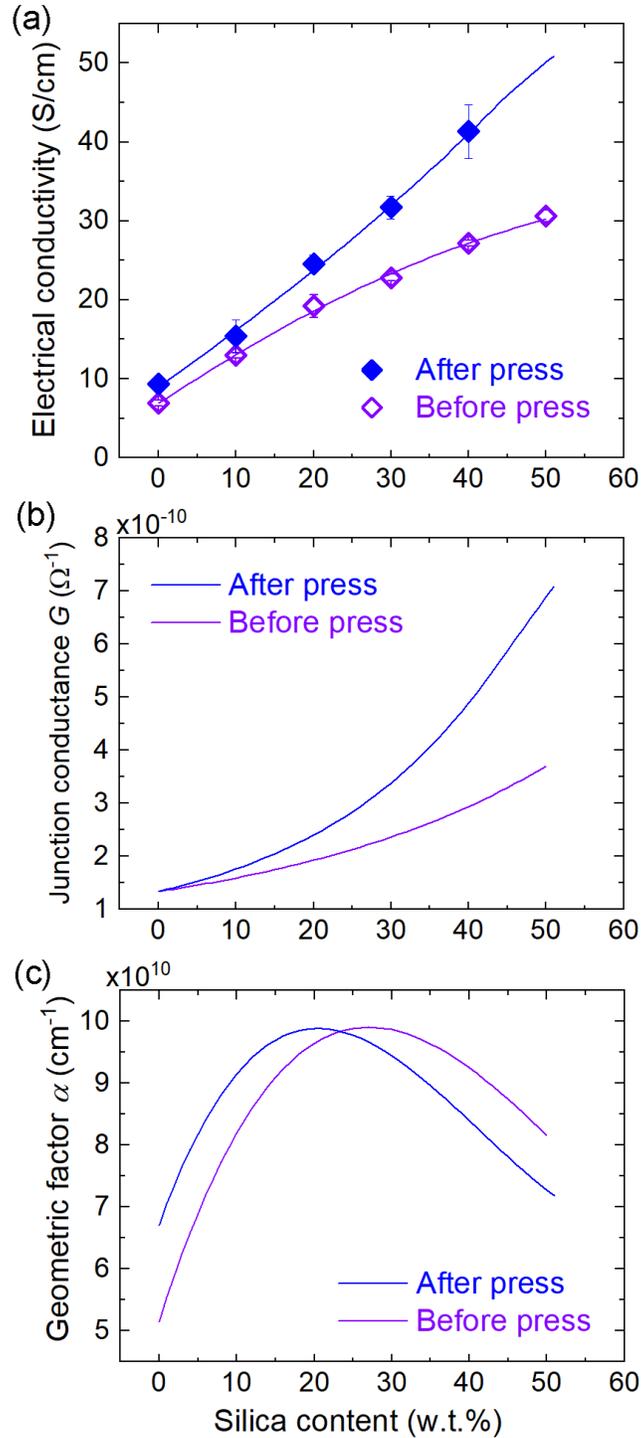

**Figure 3.** (a) Electrical conductivity of SWCNT-silica MP-PDMS composites as a function of silica w.t.% for 3 μm particle size before and after compression. The content of SWCNT was fixed at 10 w.t.%, and PDMS content was varied accordingly, i.e., decreasing PDMS w.t.% with increasing silica content. Symbols are experimental data and curves are theoretical fitting. (b)

Junction conductance $G$ calculated based on the theoretical fitting of the Seebeck coefficient shown in Fig. 2. (c) Geometric factor $\alpha$ as a function of silica content obtained from the experimental electrical conductivity in (a) and the calculated junction conductance in (b) using Eq. (1).

As shown in Figure 4, the addition of silica particles to the CNT-PDMS composite dramatically boosted the power factor. Beyond 40 w.t.% silica, the electrical conductivity slightly increases while Seebeck coefficient decreases quite much, resulting in the decrease of power factor. The maximum power factor achieved for 3 μm silica particles co-embedded in the CNT-PDMS composite are 15.48 W/mK² and 42.89 W/mK², respectively, before and after compression, which are about 6-fold and 16-fold increases, respectively.

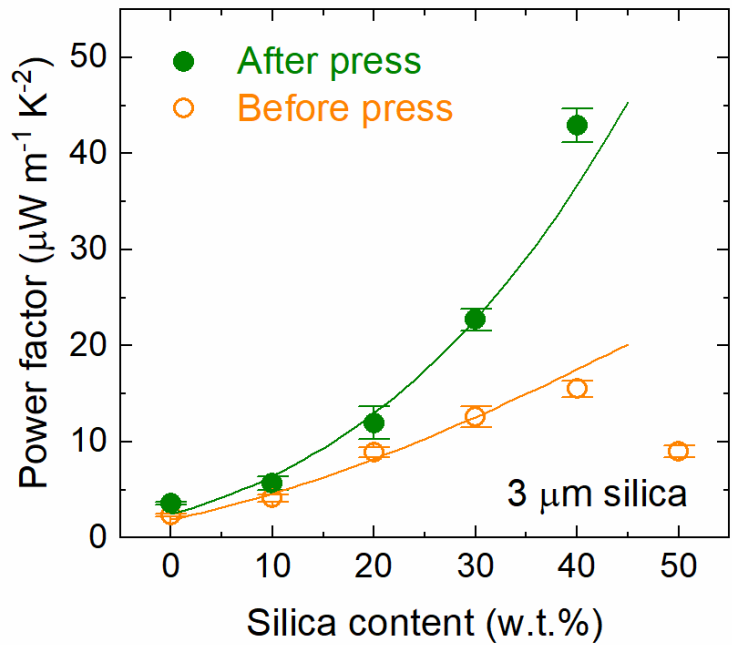

**Figure 4.** Power factor as a function of silica wt% for both 1 and 3 μm particle size, pore size – 2 nm (before and after compression), 10 %wt CNT

In addition, we obtained the Raman spectra from the samples, and the detailed results are displayed in Supporting Information: Sec. 6. Raman Spectroscopy Analysis. It is found that the G+ peak in the Raman spectra is red-shifted after sample compression, indicating longitudinal tensile strain on CNTs as a result of CNT compression by the silica microparticles. This is rarely observed in CNT-polymer composites without additional solid microparticles. The mechanical bending tests also showed that the developed CNT-PDMS composite has great mechanical stability with a maximum of ~4% reduction in electrical conductivity after being subjected to bending. See Supporting Information: Sec. 9. Bending performance for more details.

## 4. Conclusions

In short, we have demonstrated the creation of an excluded volume formed by the introduction of silica microparticles into the CNT-PDMS composite as a means to enhance the CNT networks for efficient thermoelectric energy conversion. Combined with the synergistic effect of sample compression enabled by the soft polymer matrix, this excluded volume method is useful and highly effective in boosting the power factor without any chemical treatments or CNT doping. In many cases, the power factors of various polymer-based hybrid composites were improved by increasing the electrical conductivity with little sacrifice in Seebeck coefficient. But this way, the thermal conductivity could be significantly increased due to the electronic contribution to thermal conduction, which is detrimental to the TE performance. Our excluded volume method is expected to simultaneously reduce the thermal conductivity due to the addition of highly thermally insulating silica microparticles. Further study is necessary to confirm it.

**ACKNOWLEDGMENT**

This material is based upon work supported by the National Science Foundation under Grants No. 1905571 and 1905037. V.S. and V.K.R.K. thank support from NIOSH through a grant #T42OH008432 from the Pilot Research Project Training Program of the University of Cincinnati (UC) Education and Research Center, NSF [grants IIP-2016484 and CBET-2028625], and the UC Collaborative Research Advancement Grant No. 1018371.